\documentclass[aps,prc,letterpaper,11pt,twoside,tightenlines,nofootinbib,showpacs,preprint,superscriptaddress]{revtex4-1}
\usepackage{epsfig}
\usepackage{subfigure}
\usepackage{amsmath}
\begin{document}
\arraycolsep1.5pt
\newcommand{\Ima}{\textrm{Im}}
\newcommand{\Rea}{\textrm{Re}}
\newcommand{\mev}{\textrm{ MeV}}
\newcommand{\gev}{\textrm{ GeV}}

\title{Heavy quark spin symmetric molecular states from ${\bar
    D}^{(*)}\Sigma_c^{(*)}$ and other coupled channels in the light of
the recent LHCb pentaquarks}

\author{C. W. Xiao}
\affiliation{School of Physics and Electronics, Central South University, Changsha 410083, China}

\author{J. Nieves}
\affiliation{IFIC, Centro Mixto Universitat de Val\`encia-CSIC,
Institutos de Investigaci\'on de Paterna, Aptdo. 22085, 46071 Valencia, Spain}

\author{E. Oset}
\affiliation{IFIC, Centro Mixto Universitat de Val\`encia-CSIC,
Institutos de Investigaci\'on de Paterna, Aptdo. 22085, 46071 Valencia, Spain}
\affiliation{Departamento de F\'{\i}sica Te\'orica, Universitat de Val\`encia, Spain}

\date{\today}

\begin{abstract}

We consider the ${\bar D}^{(*)}\Sigma_c^{(*)}$ states, together with
$J/\psi N$ and other coupled channels, and take an interaction
consistent with heavy quark spin symmetry, with the dynamical input
obtained from an extension of the local hidden gauge approach. By fitting only
one parameter to the recent three pentaquark states reported by the
LHCb collaboration, we can reproduce the three of them in base to the
mass and the width, providing for them the quantum numbers and
approximate molecular structure as $1/2^-$ $\bar{D} \Sigma_c$, $1/2^-$
$\bar{D}^* \Sigma_c$, and $3/2^-$ $\bar{D}^* \Sigma_c$, and isospin
$I=1/2$. We find
another state around 4374 MeV, of $3/2^-$ $\bar{D} \Sigma_c^*$
structure, for which indications appear in the experimental
spectrum. Two other near degenerate states of $1/2^-$ $\bar{D}^*
\Sigma_c^*$ and $3/2^-$ $\bar{D}^* \Sigma_c^*$ nature are also found
around 4520 MeV, which although less clear, are not incompatible with
the observed spectrum. In addition, a $5/2^-$ $\bar D^* \Sigma_c^*$
state at the same energy appears, which however does not  couple to
$J/\psi p$ in $S-$wave, and hence it is not expected to show up in the
LHCb experiment. 

\end{abstract}

\maketitle

The discovery of some pentaquarks signals by the LHCb collaboration in
2015  \cite{Aaij:2015tga,Aaij:2015fea} generated a wave of enthusiasm in the hadron physics
community.  Two states were reported, one at 4380 MeV and width
$\Gamma \sim 205\mev$ and another one at 4450 MeV and width 40
MeV. Actually there had been several predictions for hidden charm
molecular states in this region prior to the experimental discovery
\cite{Wu:2010jy,Wang:2011rga,Yang:2011wz,Yuan:2012wz,Wu:2012md,Xiao:2013yca,Uchino:2015uha,Karliner:2015ina}. The
hidden charm molecular states would have some resemblance with the
$N^*(1535)$ resonance, which in the chiral unitary approach has large
$K\Lambda$, $K\Sigma$ components
\cite{Kaiser:1995cy,Inoue:2001ip,Nieves:2001wt, Hyodo:2008xr, Gamermann:2011mq}. Large $s\bar{s}$
components in that resonance have also been claimed in
\cite{Xie:2007qt} from the study of the $pp \to pp \phi$
and $\pi^- p \to n \phi$ reactions.

A wave of theoretical papers with very different approaches,
stimulated by the LHCb findings,  were
produced trying to match the masses and
spin parity quantum numbers suggested in the experimental work,
$(3/2^-, 5/2^+)$, $(3/2^+, 5/2^-)$, $(5/2^+, 3/2^-)$ for the two
states, and other less likely combinations. In the meanwhile it has
become apparent that the hadron community took too seriously these
suggestions since the LHCb collaboration no longer sticks to any
preference for these quantum numbers \cite{tomasz}. We refer to review
papers for references to all these works
\cite{Chen:2016qju,Lebed:2016hpi,Esposito:2016noz,Guo:2017jvc,Ali:2017jda,Olsen:2017bmm,Karliner:2017qhf,Cerri:2018ypt,Liu:2019zoy}.

With the advent of Run-2 data, the LHCb collaboration updated the
results of \cite{Aaij:2015tga,Aaij:2015fea} reporting the observation
of three clear narrow structures \cite{exp3}, branded as
\begin{align}
&M_{P_{c1}} = (4311.9\pm0.7^{+6.8}_{-0.6})\, {\rm MeV}, \quad
  \Gamma_{P_{c1}}= (9.8\pm2.7^{+3.7}_{-4.5})\, {\rm MeV}, \nonumber \\
&M_{P_{c2}} = (4440.3\pm1.3^{+4.1}_{-4.7})\, {\rm MeV}, \quad
  \Gamma_{P_{c2}}= (20.6\pm4.9^{+8.7}_{-10.1})\, {\rm MeV}, \\
&M_{P_{c3}} = (4457.3\pm0.6^{+4.1}_{-1.7})\, {\rm MeV}, \quad
  \Gamma_{P_{c3}}= (6.4\pm2.0^{+5.7}_{-1.9})\, {\rm MeV}. \nonumber 
\end{align}
As one can see, the old peak at 4450 MeV is now split into two states
at 4440 MeV and 4457 MeV, the last one very narrow, and a fluctuation
observed in the old spectrum has given rise to a neat peak around 4312
MeV.

The new experimental findings have already had a reply from the
theoretical community. In \cite{Chen:2019bip} sum rules are used that
provide several scenarios to explain these states, the most favored
ones being of $\Sigma_c^{(*)}D^{(*)}$ molecular nature. In
\cite{Liu:2019tjn} heavy quark spin symmetry (HQSS) is used with
$\Sigma_c \bar{D}$, $\Sigma_c \bar{D}^*$, $\Sigma_c^* \bar{D}$,
$\Sigma_c^* \bar{D}^*$ as single channels and seven bound states are
found, three of which can be associated with the experimental
states. One should mention that in that line there is previous work,
including other coupled channels, and which also predicts seven states
with isospin $I=1/2$, and the widths of the states
\cite{Xiao:2013yca}.

Another work \cite{He:2019ify} considers again the $\Sigma_c^{(*)}D^{(*)}$
coupled channels and, using meson exchange for the dynamics, generates
three states that are associated to the new experimental resonances. There
is also an interesting suggestion to look into the isospin suppressed
$\Lambda_b \to J/\psi \Delta K^-$ reaction, showing that the ratio of
rates for $J/\psi \Delta$ to $J/\psi p$ production is largely
enhanced due to the molecular nature of the states
\cite{Guo:2019fdo}.

The blind predictions for the molecular hidden charm states have
necessarily uncertainties, which are tied to the cutoff or subtraction
constants needed to regularize the loops involved in the
calculations. The differences in the results found among different
approaches are mostly due to this point (see Refs.~\cite{Wu:2010jy}
and \cite{Wu:2012md} for instance). In this sense, differences of
masses between the $3/2^-$ and $1/2^-$ states are more reliable. Thus,
in \cite{Wu:2010jy} one finds that this difference is 149 MeV and in
\cite{Wu:2012md} it is 141 MeV. Actually these numbers are very close
to the differences between the masses of the $P_c(4457)$ and
$P_c(4312)$, which is 145 MeV. In \cite{Xiao:2013yca} this difference
is 155 MeV.

In the works of Refs.~\cite{Wu:2010jy,Wu:2012md} $\bar{D}\Sigma_c$ and
$\bar{D}^*\Sigma_c$, among other coupled channels, were used, but not
$\bar{D}\Sigma_c^*$, $\bar{D}^*\Sigma_c^*$. HQSS
\cite{IW89,hssq2,MW00} relates the strength of the interaction of
these channels and they were considered in \cite{Xiao:2013yca}. The
advent of the LHCb data offers an opportunity to tune the regulator of
the loops to adjust to some experimental data. This is the purpose of
the present work. It is similar to the study of Ref.~\cite{Liu:2019tjn}, but
includes more  channels than the $\Sigma_c^{(*)}D^{(*)}$ used
in \cite{Liu:2019tjn}, and in addition we work with coupled channels
rather than using single channels, which allows us to obtain also the
widths.

In \cite{Xiao:2013yca} the Bethe-Salpeter equation is used with the
coupled channels in $I=1/2$, $\eta_c N$, $J/\psi N$,
$\bar{D}\Lambda_c$, $\bar{D}\Sigma_c$, $\bar{D}^*\Lambda_c$,
$\bar{D}^*\Sigma_c$, $\bar{D}^*\Sigma_c^*$ for spin parity $J^P=1/2^-$
and $J/\psi N$, $\bar{D}^*\Lambda_c$, $\bar{D}^*\Sigma_c$,
$\bar{D}\Sigma_c^*$, $\bar{D}^*\Sigma_c^*$ for $J^P=3/2^-$. In
addition a single channel for $\bar{D}^*\Sigma_c^*$ in the $J^P=5/2^-$ sector
is also studied. The Bethe-Salpeter equation in matrix form 
for the scattering matrix reads
\begin{equation}
T = [1 - V \, G]^{-1}\, V,
\label{eq:BS}
\end{equation}
where $G$ is the loop function  of the meson-baryon intermediate
states and the potential $V$, respecting leading order (LO) HQSS constraints, is given
in Eqs.~\eqref{eq:ji11}--\eqref{eq:ji51} (taken from Ref.~\cite{Xiao:2013yca}).

\begin{itemize}
\item $J=1/2$, $I=1/2$
\[
\left. \phantom{(}
\begin{array}{ccccccc}
\phantom{ \sqrt{\frac{2}{3}} \text{$\mu_{13}$}} & \phantom{\frac{\sqrt{2} \text{$\mu_{13}$}}{3}} & \phantom{\sqrt{\frac{2}{3}} \text{$\mu_{23}$}} & \phantom{\frac{1}{3} \sqrt{\frac{2}{3}} (\text{$\mu_3$}-\text{$\lambda_2 $})} &
   \phantom{\frac{\sqrt{2} \text{$\mu_{23}$}}{3}} & \phantom{\frac{1}{9} \sqrt{2}
   (\text{$\mu_3$}-\text{$\lambda_2 $})} & \phantom{\frac{1}{9}
   (\text{$\lambda_2 $}+8 \text{$\mu_3$})}\\
\eta_c N & J/\psi N &  \bar D \Lambda_c &  \bar D \Sigma_c &  \bar D^* \Lambda_c
  &  \bar D^* \Sigma_c &  \bar D^* \Sigma^*_c 
\end{array}
\right. \phantom{)_{I=1/2}}
\]
\begin{equation}
\left(
\begin{array}{ccccccc}
 \text{$\mu_1$} & 0 & \frac{\text{$\mu_{12}$}}{2} &
 \frac{\text{$\mu_{13}$}}{2} & \frac{\sqrt{3} \text{$\mu_{12}$}}{2} &
 -\frac{\text{$\mu_{13}$}}{2 \sqrt{3}} & \sqrt{\frac{2}{3}}
 \text{$\mu_{13}$} \\ \\
 0 & \text{$\mu_1$} & \frac{\sqrt{3} \text{$\mu_{12}$}}{2} &
 -\frac{\text{$\mu_{13}$}}{2 \sqrt{3}} & -\frac{\text{$\mu_{12}$}}{2}
 & \frac{5 \text{$\mu_{13}$}}{6} & \frac{\sqrt{2}
   \text{$\mu_{13}$}}{3} \\ \\
 \frac{\text{$\mu_{12}$}}{2} & \frac{\sqrt{3} \text{$\mu_{12}$}}{2} &
 \text{$\mu_2$} & 0 & 0 & \frac{\text{$\mu_{23}$}}{\sqrt{3}} &
 \sqrt{\frac{2}{3}} \text{$\mu_{23}$} \\ \\
 \frac{\text{$\mu_{13}$}}{2} & -\frac{\text{$\mu_{13}$}}{2 \sqrt{3}} & 0 & \frac{1}{3} (2 \text{$\lambda_2 $}+\text{$\mu_3$}) & \frac{\text{$\mu_{23}$}}{\sqrt{3}} & \frac{2 (\text{$\lambda_2$}-\text{$\mu_3$})}{3 \sqrt{3}} & \frac{1}{3} \sqrt{\frac{2}{3}}
 (\text{$\mu_3$}-\text{$\lambda_2 $}) \\ \\
 \frac{\sqrt{3} \text{$\mu_{12}$}}{2} & -\frac{\text{$\mu_{12}$}}{2} & 0 & \frac{\text{$\mu_{23}$}}{\sqrt{3}} & \text{$\mu_2$} & -\frac{2 \text{$\mu_{23}$}}{3} & \frac{\sqrt{2} \text{$\mu_{23}$}}{3} \\ \\
 -\frac{\text{$\mu_{13}$}}{2 \sqrt{3}} & \frac{5 \text{$\mu_{13}$}}{6} & \frac{\text{$\mu_{23}$}}{\sqrt{3}} & \frac{2 (\text{$\lambda_2 $}-\text{$\mu_3$})}{3 \sqrt{3}} & -\frac{2 \text{$\mu_{23}$}}{3} & \frac{1}{9} (2 \text{$\lambda_2 $}+7 \text{$\mu_3$}) &
 \frac{1}{9} \sqrt{2} (\text{$\mu_3$}-\text{$\lambda_2 $}) \\ \\
 \sqrt{\frac{2}{3}} \text{$\mu_{13}$ } & \frac{\sqrt{2} \text{$\mu_{13}$}}{3}\; & \sqrt{\frac{2}{3}} \text{$\mu_{23}$ } & \frac{1}{3} \sqrt{\frac{2}{3}} (\text{$\mu_3$}-\text{$\lambda_2 $})\; &
   \frac{\sqrt{2} \text{$\mu_{23}$}}{3}\;\; & \frac{1}{9} \sqrt{2}
   (\text{$\mu_3$}-\text{$\lambda_2 $}) & \frac{1}{9}
   (\text{$\lambda_2 $}+8 \text{$\mu_3$}) 
\end{array}
\right)_{ I=1/2}
\label{eq:ji11}
\end{equation}
\newpage

\item  $J=3/2$, $I=1/2$  
\[
\left. \phantom{(}
\begin{array}{ccccc}
\phantom{\frac{\sqrt{5} \text{$\mu_{13}$}}{3}} & \phantom{\frac{\sqrt{5}
   \text{$\mu_{23}$}}{3}} & \phantom{\frac{1}{9} \sqrt{5}
 (\text{$\mu_3$}-\text{$\lambda_2 $})} & \phantom{\frac{1}{3} \sqrt{\frac{5}{3}} 
(\text{$\lambda_2$}-\text{$\mu_3$})} & \phantom{\frac{1}{9} (4 \text{$\lambda_2 $}+5 \text{$\mu_3$})}\\
 J/\psi N &  \bar D^* \Lambda_c &  \bar D^* \Sigma_c 
&  \bar D \Sigma^*_c  &  \bar D^* \Sigma^*_c 
\end{array}
\right. \phantom{)_{I=1/2}}
\]
\begin{equation}
\left(
\begin{array}{ccccc}
 \text{$\mu_1$} & \text{$\mu_{12}$} & \frac{\text{$\mu_{13}$}}{3} & -\frac{\text{$\mu_{13}$}}{\sqrt{3}} & \frac{\sqrt{5} \text{$\mu_{13}$}}{3} \\\\
 \text{$\mu_{12}$} & \text{$\mu_2$} & \frac{\text{$\mu_{23}$}}{3} & -\frac{\text{$\mu_{23}$}}{\sqrt{3}} & \frac{\sqrt{5} \text{$\mu_{23}$}}{3} \\\\
 \frac{\text{$\mu_{13}$}}{3} & \frac{\text{$\mu_{23}$}}{3} & \frac{1}{9} (8 \text{$\lambda_2 $}+\text{$\mu_3$}) & \frac{\text{$\lambda_2 $}-\text{$\mu_3$}}{3 \sqrt{3}} & \frac{1}{9}
   \sqrt{5} (\text{$\mu_3$}-\text{$\lambda_2 $}) \\\\
 -\frac{\text{$\mu_{13}$}}{\sqrt{3}} & -\frac{\text{$\mu_{23}$}}{\sqrt{3}} & \frac{\text{$\lambda_2 $}-\text{$\mu_3$}}{3 \sqrt{3}} & \frac{1}{3} (2 \text{$\lambda_2 $}+\text{$\mu_3$}) &
   \frac{1}{3} \sqrt{\frac{5}{3}} (\text{$\lambda_2 $}-\text{$\mu_3$}) \\\\
 \frac{\sqrt{5} \text{$\mu_{13}$}}{3}\; & \frac{\sqrt{5}
   \text{$\mu_{23}$}}{3}\; & \frac{1}{9} \sqrt{5}
 (\text{$\mu_3$}-\text{$\lambda_2 $})\; & \frac{1}{3} \sqrt{\frac{5}{3}} 
(\text{$\lambda_2$}-\text{$\mu_3$})\; & \frac{1}{9} (4 \text{$\lambda_2 $}+5 \text{$\mu_3$}) \\
\end{array}
\right)_{I=1/2}
\label{eq:ji31}
\end{equation}
\item $J=5/2$, $I=1/2$
  \begin{equation}
    {\bar D}^* \Sigma_c^*: \, (\lambda_2)_{I=1/2}
    \label{eq:ji51}
    \end{equation}
\end{itemize}
LO HQSS interactions for $I=3/2$ can be also found in
Ref.~\cite{Xiao:2013yca}.

Note, that the single channel
interactions used in \cite{Liu:2019tjn} are recovered from
Eqs.~\eqref{eq:ji11}--\eqref{eq:ji51}, identifying the terms $C_a$ and
$C_b$ introduced in that reference to $(2\lambda_2/3+\mu_3/3)_{I=1/2}$
and $(\lambda_2/3-\mu_3/3)_{I=1/2}$, respectively.

There are seven parameters relying upon
HQSS only, but when one imposes a particular dynamics, restrictions
among them appear, as shown in \cite{Garcia-Recio:2013gaa}. In the
present work we shall consider the same constraints as in
\cite{Xiao:2013yca}, which stem from the use of an extension of the
local hidden gauge approach, where the source of interaction is the
exchange of vector mesons
\cite{Bando:1984ej,Bando:1987br,Meissner:1987ge}. Detailed discussions
justifying this extension to the charm, or bottom sector, are given in
\cite{Sakai:2017avl,Debastiani:2017ewu}. These constraints are for
$I=1/2$
\begin{equation}
\begin{split}
\mu_1 &= 0, \quad \mu_{23} = 0, \quad \lambda_2 = \mu_3, \quad \mu_{13} = -\mu_{12}, \\
\mu_2 &= \frac{1}{4f^2} (k^0 + k'^0),\quad \mu_3 = -\frac{1}{4f^2} (k^0 + k'^0),\\
\mu_{12} &= -\sqrt{6}\ \frac{m_\rho^2}{p^2_{D^*} - m^2_{D^*}}\ \frac{1}{4f^2}\ (k^0 + k'^0),\label{eq:ji11fi}
\end{split}
\end{equation} 
with $f_\pi = 93 \mev$, and $k^0$, $k'^0$ the center of mass energies
of the mesons in the $MB \to M' B'$ transition. In addition,
$p^2_{D^*}$ applies to the $t-$channel exchanged $D^*$ in the tree
level of some suppressed transitions ($\eta_c N\to \bar{D}\Lambda_c$
for instance).

The novelty with respect to Ref. \cite{Xiao:2013yca} is a different
choice of the subtraction constant to renormalize the meson-baryon
loops ($G$) in dimensional regularization. A subtraction constant $a(\mu)=-2.3$ with
$\mu=1$ GeV was used in \cite{Xiao:2013yca}. This value was
justified since it falls in the range of ``natural values'' discussed
in \cite{Oller:2000fj} and was also used in \cite{Wu:2010jy}. The
scheme produces seven states, three of which  can be clearly associated to the
recently found experimental resonances. The new information allows us to
take a new value of $a(\mu=1\,{\rm GeV})=-2.09$, such that the sum of masses of the
three theoretical states matches the experimental results.  With this
constraint we fix the only free parameter of the model of
Ref. \cite{Xiao:2013yca}. The results are reported in Tables
\ref{tab:cou11} and \ref{tab:cou31} for $J^P= 1/2^-$ and $3/2^-$, respectively. In addition we get a mass of
4519.23 MeV and a zero width for the single channel $\bar{D}^*
\Sigma_c^*$ with $J=5/2^-$. This channel obviously does not couple to
$J/\psi N$ so we should not see it in the $\Lambda_b \to J/\psi p K^-$
experiment. The states in Tables \ref{tab:cou11} and \ref{tab:cou31} all
couple to $J/\psi N$ and in principle they could be seen in the
experiment, although we cannot predict their strength in the spectrum. In
Table. \ref{tab:sumre}, we show the results for the three resonances that
we identify with the experimental states. The main channel is taken from
the largest coupling.  We find the last two states nearly degenerate,
yet, the widths of the states force us to identify the $3/2^-$ state
with the $P_c(4457)$. Note that the masses divert only in a few MeV
from the experimental ones, and the three widths obtained are
compatible with the experiment. The results of Table \ref{tab:sumre}
are similar to those of \cite{Liu:2019tjn}, where the input has been
adjusted to reproduce the $P_c(4440)$ and $P_c(4457)$ states. There is
only a small difference since in \cite{Liu:2019tjn} the $J^P$
assignments to the $P_c(4440)$ and $P_c(4457)$ are opposite to
ours. Our approach, providing the width, gives us one additional
reason to support our assignment. As to the molecular nature of the
states, the single channel calculation of \cite{Liu:2019tjn} gives the
same state as those written in Table \ref{tab:sumre} as our main
channel.

\begin{table}[ht]
     \renewcommand{\arraystretch}{1.2}
\centering
\caption{Dimensionless coupling constants of the $(I=1/2, J^P=1/2^-)$ poles found in this
  work to the different channels.} \label{tab:cou11}
\begin{tabular}{cccc cccc}
\hline\hline
\multicolumn{2}{c}{$(4306.38+i7.62)$ MeV}  \\
\hline
   & $\eta_c N$ & $J/\psi N$ & $\bar{D} \Lambda_c$ & $\bar{D} \Sigma_c$ & $\bar{D}^* \Lambda_c$ & $\bar{D}^* \Sigma_c$ & $\bar{D}^* \Sigma_c^*$  \\
\hline
$g_i$ & $0.67+i0.01$ & $0.46-i0.03$ & $0.01-i0.01$
& $\mathbf{2.07-i0.28}$ & $0.03+i0.25$ & $0.06-i0.31$ & $0.04-i0.15$  \\
$|g_i|$ & $0.67$ & $0.46$ & $0.01$ & $2.09$ & $0.25$ & $0.31$ & $0.16$  \\
\hline
\multicolumn{2}{c}{($4452.96+i11.72$) MeV}  \\
\hline
    & $\eta_c N$ & $J/\psi N$ & $\bar{D} \Lambda_c$ & $\bar{D} \Sigma_c$ & $\bar{D}^* \Lambda_c$ & $\bar{D}^* \Sigma_c$ & $\bar{D}^* \Sigma_c^*$  \\
\hline
$g_i$ & $0.24+i0.03$ & $0.88-0.11$ & $0.09-i0.06$ & $0.12-i0.02$ &
$0.11-i0.09$ &
$\mathbf{1.97-i0.52}$ & $0.02+i0.19$  \\
$|g_i|$ & $0.25$ & $0.89$ & $0.11$ & $0.13$ & $0.14$ & $2.03$ & $0.19$  \\
\hline
\multicolumn{2}{c}{$(4520.45+i11.12)$ MeV}  \\
\hline
   & $\eta_c N$ & $J/\psi N$ & $\bar{D} \Lambda_c$ & $\bar{D} \Sigma_c$ & $\bar{D}^* \Lambda_c$ & $\bar{D}^* \Sigma_c$ & $\bar{D}^* \Sigma_c^*$  \\
\hline
$g_i$ & $0.72-i0.10$ & $0.45-i0.04$ & $0.11-i0.06$ & $0.06-i0.02$ & $0.06-i0.05$ & $0.07-i0.02$ & $\mathbf{1.84-i0.56}$  \\
$|g_i|$ & $0.73$ & $0.45$ & $0.13$ & $0.06$ & $0.08$ & $0.08$ & $1.92$  \\
\hline
\end{tabular}
\end{table}

\begin{table}[ht]
     \renewcommand{\arraystretch}{1.2}
\centering
\caption{Same as Table~\ref{tab:cou11} for $J^P=3/2^-$.} \label{tab:cou31}
\begin{tabular}{ccc ccc}
\hline\hline
$(4374.33+i6.87)$ MeV & $J/\psi N$ & $\bar{D}^* \Lambda_c$ & $\bar{D}^* \Sigma_c$ & $\bar{D} \Sigma_c^*$ & $\bar{D}^* \Sigma_c^*$  \\
\hline
$g_i$ & $0.73-i0.06$ & $0.11-i0.13$ & $0.02-i0.19$ & $\mathbf{1.91-i0.31}$ & $0.03-i0.30$  \\
$|g_i|$ & $0.73$ & $0.18$ & $0.19$ & $1.94$ & $0.30$  \\
\hline
$(4452.48+i1.49)$ MeV & $J/\psi N$ & $\bar{D}^* \Lambda_c$ & $\bar{D}^* \Sigma_c$ & $\bar{D} \Sigma_c^*$ & $\bar{D}^* \Sigma_c^*$  \\
\hline
$g_i$ & $0.30-i0.01$ & $0.05-i0.04$ & $\mathbf{1.82-i0.08}$ & $0.08-i0.02$ & $0.01-i0.19$  \\
$|g_i|$ & $0.30$ & $0.07$ & $1.82$ & $0.08$ & $0.19$   \\
\hline
$(4519.01+i6.86)$ MeV & $J/\psi N$ & $\bar{D}^* \Lambda_c$ & $\bar{D}^* \Sigma_c$ & $\bar{D} \Sigma_c^*$ & $\bar{D}^* \Sigma_c^*$  \\
\hline
$g_i$ & $0.66-i0.01$ & $0.11-i0.07$ & $0.10-i0.3$ & $0.13-i0.02$ & $\mathbf{1.79-i0.36}$  \\
$|g_i|$ & $0.66$ & $0.13$ & $0.10$ & $0.13$ & $1.82$   \\
\hline
\end{tabular}
\end{table}

\begin{table}[htb]
\renewcommand{\arraystretch}{1.7}
     \setlength{\tabcolsep}{0.2cm}
\centering
\caption{Identification of some of the $I=1/2$ resonances found in
  this work  with experimental states.}
\label{tab:sumre}
\begin{tabular}{lccc cc}
\hline
Mass [MeV] & Width [MeV] & Main channel & $J^P$   &  Experimental state   \\
\hline\hline
4306.4 & 15.2 & $\bar D \Sigma_c$      &  $1/2^-$ & $P_c(4312)$   \\
\hline
4453.0 & 23.4 & $\bar D^* \Sigma_c$   & $1/2^-$ &  $P_c(4440)$  \\
\hline
4452.5 & 3.0    & $\bar D^* \Sigma_c$   & $3/2^-$ &  $P_c(4457)$  \\
\hline
\end{tabular}
\end{table}

We should note that the reason why $\mu_{23}=0$ in
Eq. \eqref{eq:ji11fi} is the neglect of pion exchange which was found
small, although not negligible in \cite{Xiao:2013yca}. Its
consideration would break the near degeneracy that we have in the two
higher states of Table \ref{tab:sumre}, as was found in
\cite{Uchino:2015uha}, where, however, the effect of pion exchange was
found more important as a consequence of the choice of large cutoffs
that made the binding much larger.

It looks strange that the widths obtained here are smaller than those
reported in \cite{Xiao:2013yca} in spite that the masses of the
states are bigger and, hence, there is more phase space for decay. The
answer has to be found in the fact that the couplings have also become
smaller. This is not an accident but the consequence of one important
property. Indeed it is well known that in the case of a one channel
bound state, the coupling square, $g^2$, goes as the square root of
the binding energy as a consequence of the most celebrated Weinberg's
compositeness condition \cite{Weinberg:1965zz,Baru:2003qq}. It is,
however, less known that in the case of coupled channels, all
couplings go to zero close to the threshold of one channel
\cite{toki,Gamermann:2009uq}. In the present case the $P_c(4312)$ is
close to the $\Sigma_c\bar{D}$ threshold and the $P_c(4440)$ and
$P_c(4457)$ are very close to the $\Sigma_c\bar{D}^*$ threshold.

The association that we have done of the states found in this work
with the experimental ones agrees with the one proposed in
\cite{He:2019ify} where, however, the widths are not evaluated. One
should also note that in \cite{Liu:2019tjn} and here we find seven
states, while only three states are reported in
\cite{He:2019ify}. Actually, it is worth noting that in
\cite{Liu:2019tjn} a $3/2^-$ $\bar{D}\Sigma_c^*$ state is reported at
4371 MeV, while we find a state in Table \ref{tab:cou31}, coupling
mostly to $\bar{D}\Sigma_c^*$, at 4374 MeV with a width of about 14
MeV. It is interesting to call the attention to the fact that the
$J/\psi p$ spectrum of Ref.~\cite{exp3} shows a peak around 4370 MeV
that could have well been identified with a new state.  The strength
of this peak is only about 1/2 of that of the $P_c(4312)$ and it is
clearly distinguishable from other minor peaks that can be consistent
with statistical fluctuations. We find two more states that can decay
to $J/\psi p$ in Tables \ref{tab:cou11} and \ref{tab:cou31}, a state
of $1/2^-$ at 4520 MeV and a $3/2^-$ state at 4519 MeV, which couple
mostly to $\bar{D}^*\Sigma_c^*$. The single channel results reported
in \cite{Liu:2019tjn} also find these two states at 4523 MeV and 4517
MeV, respectively, in their option A. With the risk of stretching too
much the imagination there is indeed a peak in $J/\psi p$ spectrum of
\cite{exp3} in that region that, however, it could as well be a
statistical fluctuation. Note that we also obtain a near degenerate
state with this nature for $5/2^-$. This state appears at 4500 MeV in
option A and at 4523 in option B of \cite{Liu:2019tjn}.

In summary, the molecular picture in the coupled channels to
$J/\psi p$ in S-wave, using constraints of HQSS and dynamics from the
extension of the local hidden gauge approach, basically an extension
of the chiral unitary approach to the charm sector, renders six states
that couple to $J/\psi p$. Three of these resonances can be identified with the
three states reported in \cite{exp3} in base to their masses and
widths. In addition,  we  provide a prediction of
their $J^P$ quantum numbers and of the nature
of these states as basically $1/2^-$ $\bar{D}\Sigma_c$, $1/2^-$
$\bar{D}^*\Sigma_c$ and $3/2^-$ $\bar{D}^*\Sigma_c$. We find a fourth
state, which couples mostly to $\bar{D}\Sigma_c^*$ with $3/2^-$, for
which there are indications in the $J/\psi p$ spectrum of
\cite{exp3}. The other two states, of $\bar D^*  \Sigma^*_c$ nature,
are around 4520 MeV (close to the threshold of this meson-baryon pair), and although there are small peaks in that region
in \cite{exp3}, one can only speculate at the present stage.
They are also near degenerate with a $5/2^-$
state of the same nature, which however is not expected to show up in
the LHCb experiment. This degeneracy is obvious from the diagonal
$\bar{D}^*\Sigma_c^*$  interactions given in
Eqs.~\eqref{eq:ji11}--\eqref{eq:ji51}, taking into account that the
hidden gauge model used here leads to $\lambda_2=\mu_3$ for $I=1/2$.

\section*{Acknowledgments}
This research  has been supported by the Spanish Ministerio de
Ciencia, Innovaci\'on  y Universidades and European FEDER funds under
Contracts FIS2017-84038-C2-1-P,  FIS2017-84038-C2-2-P and SEV-2014-0398.

\end{document}